\documentstyle [12pt]{article}

\begin{document}
\setcounter{page}{1}
\title{(Super)rare decays of an extra $Z^{\prime}$ boson via Higgs
boson emission}
\author{G.A. Kozlov\\
\em Bogolyubov Laboratory of Theoretical Physics\\
\em Joint Institute for Nuclear Research\\
\em 141980 Dubna, Moscow Region, Russia}
\date{}
\maketitle
\begin{abstract}

{\small The phenomenological model of an extra U(1) neutral gauge
$Z^{\prime}$ boson coupled to heavy quarks is presented. In particular,
we discuss the probability for a light $Z_{2}$ mass eigenstate decay into
a bound state composed of heavy quarks via Higgs boson emission.\\

 }
\end{abstract}

1. Theoretical interest in an extra neutral vector boson $Z^{\prime}$ is
mainly motivated by experimental observation of possible deviations from
the Standard Model (SM) predictions for the decay of the SM $Z$ boson into
$\bar{c}c$- and $\bar{b}b$- pairs of quarks ($R_c$- and $R_b$-ratios) [1].
The deviations may be considered as one of the indications of new physics
(NP) beyond the SM. The promising explanation of the observed phenomena
is implied in the extra $Z^{\prime}$ models (see refs. [7-16] in [2]).
New gauge bosons can be detected in future high-energy colliders, namely,
Large Hadron Collider (LHC) at CERN, which can test the nature and structure
of many theoretical models at a scale of 1 TeV, at least. Theoretical
predictions of new neutral or charged gauge bosons come from various
extensions of the SM [3]. New extra bosons naturally appear in the Grand
Unification Theory (GUT) models [3]. A simple and well-known version
among the extensions of the SM is the minimal one, aimed at unifying
interactions, the $E_{6}$ GUT model [3]. Since the
breaking of $E_{6}$ GUT into SM is accompanied by at least one extra U(1)
group
$(E_{6}\rightarrow SU(3)_{C}\times SU(2)_{L}\times U(1)_{Y}\times U(1))$,
there may exist a heavy neutral boson $Z^{\prime}$ which can mix with
an ordinary $Z$ boson.
There are two new gauge bosons appearing in
$E_{6}$ GUT models [3] where only one originates from the SO(10) subgroup
$$ E_{6}\supset SO(10)\times U(1)_{\Psi}\ ,$$
$$ SO(10)\supset SU(5)\times U(1)_{\chi}\ ,$$
$$ SU(5)\supset SU(3)_{C}\times SU(2)_{L}\times U(1)_{Y} ,$$
while the $Z^{\prime}$ boson is a composition of $Z_{\Psi}$- and $Z_{\chi}$-
components mixed with a free angle $\Theta$ [3]:
$$ Z^{\prime}=Z_{\Psi}\ cos{\Theta}-Z_{\chi}\ sin{\Theta}\ .$$

 The best sensitivity to a possible signal from $Z^{\prime}$ is
achieved through the decay channel, $Z^{\prime}\rightarrow\bar{Q}Q$,
where $Q$ ($\bar{Q}$) stands for a heavy quark (antiquark).
The decay $Z^{\prime}\rightarrow\bar{Q}Q$ is the most important though not the
only production signal possible for $Z^{\prime}$.
To search for $Z^{\prime}$ at LHC, it is important to know as much as
possible about their decay modes in both the standard Drell-Yan (DY)-
type sectors and the (super)rare ones. Other
channels can provide important information on the $Z^{\prime}$ boson
couplings. If we go beyond SM, there are several possibilities for some
quark bound state resonances $B\equiv\{\bar{Q}Q\}$ to be produced
via the particle
interplay accompanied by the Higgs-boson ($H$) emission. A possible extension
of the SM adopting a $Z^{\prime}$ boson may need more unknown
$h$-fermions (spin-1/2 heavy exotic quarks) and $H$ particles to be included.
It is known
that the $Z$ boson is not yet an exact mass eigenstate, but turns out to be
 mixed with $Z^{\prime}$. In the $Z-Z^{\prime}$ mixing scheme the mass
eigenstates $Z_{1}$ and $Z_{2}$ are rotated with respect to the basis $Z$
and $Z^{\prime}$
$$\pmatrix{Z_{1}\cr
Z_{2}\cr}=\pmatrix{\cos\kappa&\sin\kappa\cr
-\sin\kappa&\cos\kappa\cr}\pmatrix{Z\cr
Z^{\prime}\cr}\ ,$$
by the mixing angle $\kappa$
$$\kappa=\arctan\left (\frac{M_{Z}^2-M_{Z_{1}}^2}{M_{Z_{2}}^2-M_{Z}^2}
\right )^{1/2} $$
with $M_{Z_{1}}$ and $M_{Z_{2}}$ being the masses for mass eigenstates
$Z_{1}$ and $Z_{2}$, respectively.

In this letter, we study a possible extra $Z_{2}$ state and its
interpretations which have direct implications for NP at LHC. Our
interest is in $Z_{2}$ production and possible pair production process
$Z_{2}(W,Z)$ with $Z_{2}$ decay into pairs of heavy quarks leading to
$\bar{Q}Q$ and $\bar{Q}Q(W,Z)$ events at LHC with $\bar{Q}Q$ invariant
mass peaked at the mass up to an order of O(0.4 TeV). If the
$Z_{2}$ state is heavy enough to produce the $H$ boson, one can determine the
effective coupling of the $Z_{2}$-$H$ interaction. We are to give the
estimation of the partial decay widths $\Gamma$ ratio
\begin{eqnarray}
\label{e1}
R(Z_{2}\rightarrow H \{\bar{Q}Q\}_{s=1}/\bar{Q}Q)\equiv
\frac{\Gamma(Z_{2}\rightarrow H \{\bar{Q}Q\}_{s=1})}
{\Gamma(Z_{2}\rightarrow\bar{Q}Q)}\ ,
\end{eqnarray}
where $\{\bar{Q}Q\}_{s=1}$ stands for a spin-1 quark-antiquark bound state.

2. To analyze the $Z_{2}$ state effects within the model under consideration,
let us concentrate on the $Z_{2}$ couplings. Neglecting the interactions of
$Z$ bosons to leptons ("leptophobic" character of $Z$ bosons) the interactions
of mass eigenstates $Z_{i}$ ($i>1$) with heavy quarks are described by the
following Lagrangian density:
\begin{eqnarray}
\label{e2}
-L_{Z_{i}Q}=g_{Z}\sum^{\infty}_{i=1}\sum_{Q}\bar{Q}(g_{V_{i}}-g_{A_{i}}
\gamma_{5})\gamma^{\mu}Q Z_{i_{\mu}}\ ,
\end{eqnarray}
where one of the sums runs over all heavy quarks $Q$, $g_{Z}$ is presented as
the SM coupling $g/\sqrt{1-s^{2}_{W}}$ ($s_{W}\equiv\sin\Theta_{W}$),
$Z_{1_{\mu}}$ is understood as the SM $Z$ boson field while $Z_{j}$
with $j\geq$ 2 are
additional $Z$ states in the weak-eigenstate basis. We shall consider the
model with one light $Z_{2}$ mass eigenstate only. The vector
$g_{V_{i}}$ and the axial
$g_{A_{i}}$ couplings ($i$=1,2) in (\ref{e2}) are defined as
\begin{eqnarray}
\label{e3}
 g_{V_{1}}=g_{V}\cos\kappa +g^{\prime}_{V}\alpha\sin\kappa\ \ ,
 g_{A_{1}}=g_{A}\cos\kappa  +g^{\prime}_{A}\alpha\sin\kappa\ ,
\end{eqnarray}
\begin{eqnarray}
\label{e4}
g_{V_{2}}=\alpha g_{V}^{\prime}\cos\kappa-g_{V}\sin\kappa\ \ ,
g_{A_{2}}=\alpha g^{\prime}_{A}\cos\kappa-g_{A}\sin\kappa
\end{eqnarray}
with
$$g_{V}=\frac{1}{2}T_{3L}-s^{2}_{W}\cdot e_{Q}\ ,\
g_{A}=\frac{1}{2}T_{3L}\ ,$$
for $T_{3L}$ and $e_{Q}$ being the third component of the weak isospin and
the electric charge, respectively. Both $g^{\prime}_{V}$ and $g^{\prime}_{A}$
in (\ref{e3})-(\ref{e4})
represent the chiral properties of the $Z^{\prime}$ boson interplay with
quarks and the relative strengths of these interactions
\begin{eqnarray}
\label{e5}
-L_{Z^{\prime}Q}=g^{\prime}_{Z}\sum_{Q}\bar{Q}(g^{\prime}_{V}-
g^{\prime}_{A}\gamma_{5})\gamma^{\mu}Q Z^{\prime}_{\mu}\ .
\end{eqnarray}
For the GUT models, the free parameter $g^{\prime}_{Z}$ in (\ref{e5}) is
related to $\alpha$ in (\ref{e3})-(\ref{e4}) as
$\alpha\equiv (g^{\prime}_{Z}/g_{Z})=\sqrt{(5/3)\omega}\cdot s_{W}$ [4],
where $\omega$ depends on the symmetry breaking pattern and the fermion
sector of the model, but is usually taken $\omega\sim$ 2/3-1. The choice
of $\alpha\simeq$ 0.62 provides the equality of both $g_{Z}$ and
$g^{\prime}_{Z}$ on the scale of the mass of the unification $M_{X}\simeq
M_{GUT}$ into $E_{6}$. Neglecting some differences in the renormalization
group evolution of both $g_{Z}$ and $g^{\prime}_{Z}$, one can deal with
$\alpha$ at the energies $\sim M_{Z^{\prime}}\sim M_{Z_{2}}\sim$ O(1 TeV).

Suppose that the $Z_{2}$ state could be produced at LHC via $\bar{Q}Q
\rightarrow Z_{2}$ subprocess, and in the narrow $Z_{2}$ width approximation
 the cross section
$$\sigma (\bar{Q}Q\rightarrow Z_{2})=K(M_{Z_{2}})\frac{2\pi G_{F}
M^{2}_{Z_{1}}}
{3\ \sqrt{2}}(g^{2}_{V_{2}}+g^{2}_{A_{2}})\delta (s-M_{Z_{2}}) $$
is both $M_{Z_{2}}$- and $\kappa$- dependent. Here, $G_{F}$ is the Fermi
constant and $K$ factor reflects the higher order QCD process [5]
$$ K(M_{Z_{2}})=1+\frac{\alpha_{s}(M^{2}_{Z_{2}})}{2\ \pi}\frac{4}{3}\left (
1+\frac{4}{3}\ \pi^2\right )\ .$$
Note that two-loop $\alpha_{s}(M^{2}_{Z_{2}})\sim $0.1 for $\Lambda_{QCD}$=
200 MeV at $M_{Z_{2}}< 2 m_{t}$ (5 flavors) and $ M_{Z_{2}}> 2 m_{t}$
(6 flavors) for $m_{t}$ being the top quark mass [2].

The partial width for $Z_{2}$ decays into quarks is determined by the
couplings $g_{V_{2}}$ and $g_{A_{2}}$ (\ref{e4}), namely (the number
of colors $N_{c}$=3 is taken into account)
\begin{eqnarray}
\label{e6}
\Gamma(Z_{2}\rightarrow\bar{Q}Q)
=\frac{2\ G_{F}\ M^{2}_{Z}}{\sqrt{2}\ \pi}\ C(M^{2}_{Z_{2}})\ M_{Z_{2}} \cr
\times {(1-4r_{q})}^{1/2}
\left [g^{2}_{V_{2}}(1+2r_{q})+g^{2}_{A_{2}}(1-4r_{q})\right ].
\end{eqnarray}
Here, $r_{q}\equiv m^{2}/M^{2}_{Z_{2}}$, $M_{Z}$ and $m$ are the masses of
the $Z$ boson and a quark, respectively, while the $C$ factor is defined
by the running strong coupling constant $\alpha_{s}$
$$C(\mu^2)=1+\frac{\alpha_{s}(\mu^2)}{\pi}+1.409\frac{\alpha^{2}_{s}
(\mu^2)}{\pi^{2}}-12.77\frac{\alpha^{3}_{s}(\mu^2)}{\pi^{3}}\ $$
with an arbitrary scale $\mu$. The interactions of the $Z_{2}$ state with
quarks are expressed in terms of three parameters $x$, $y^{u}$ and $y^{d}$
[1], where labels $u$ and $d$ mean the up- and down-type of quarks
$$ 2g^{u}_{V_{2}}=x+y^{u}\ \ \ ,-2g^{u}_{A_{2}}=-x+y^{u}\ \ ,$$
$$ 2g^{d}_{V_{2}}=x+y^{d}\ \ \ ,-2g^{d}_{A_{2}}=-x+y^{d}\ \ ,$$

3. The $\{\bar{Q}Q\}_{s=1}$ bound state with the 4-momentum $Q_\mu$ and the
mass $m_{B}$ may be produced in the $Z_{2}$ state decay via the $H$ boson
emission
 with the 4-momentum $k_{\mu}$ in a heavy quark-loop scheme.
The decay amplitude is written as
\begin{eqnarray}
\label{e7}
 A(k,Q)=\int d_{4}q\ Tr\left\{\Gamma^{+}_{Q}(q)\cdot\sum^{3}_{i=1}T_{i}
(q,k;Q)\right\}\ ,
\end{eqnarray}
where $\Gamma_{Q}(q_{\mu})$ is the spin-1 quark bound state vertex function
depending on the relative momentum $q_{\mu}$ of $\bar{Q}$ and $Q$,
 while $T_{i}$ are the rest of the total matrix element. In fact, $T_{i}$
in (\ref{e7})
carry the dependence of interplay of $H$ to heavy quarks ($i$=1,2) and
$Z_{2}$ state to $H$ boson ($i$=3) via the couplings $g_{H}=m/{\langle
H\rangle}_{0}$ and $g_{Z_{2}H}=2M^{2}_{Z_{2}}/{\langle H\rangle}_{0}$,
respectively, where ${\langle H\rangle}_{0}$ stands for the vacuum expectation
value of the singlet field $H$.
 Generally, $\Gamma_{Q}(q_{\mu})$ is constructed [6]
in terms of quark $u(Q_{\mu})$ and antiquark $v(\bar{Q}_{\mu})$ spinors in a
given spin configuration accompanied by the covariant confinement-type
wave function $\phi_{Q} (q^2;\beta)$ (in $\Re(S_{4})$) containing the
model parameter $\beta$ [7]
\begin {eqnarray}
\label{e8}
\Gamma_{Q}(q_{\mu})=\bar{u}(Q_{\mu})\frac{\delta^{j}_{i}}{\sqrt{3}}\
U_{\alpha\beta}\ \phi_{Q}(q^2;\beta)\ v(\bar{Q}_{\mu})\ .
\end{eqnarray}
Here, the second rank symmetric spinors $U_{\alpha\beta}$
obey the standard Bargman-Wigner equations [8] $\left (\not Q-m_{B}
\right )^{\alpha^{\prime}}_{\alpha}U_{\alpha^{\prime}\beta}=0$.

The decay width of $Z_{2}\rightarrow H\{\bar{Q}Q\}_{s=1}$ is given by
\begin{eqnarray}
\label{e9}
\Gamma(Z_{2}\rightarrow H\{\bar{Q}Q\}_{s=1})=
\frac{g^{2}_{Z}g^{2}_{V_{2}}M^{3}_{Z_{2}}N_{C}\cos^{2}{\vartheta}
x^{2}_{\beta}\sqrt{\lambda (1,x_{H},x_{B})}}
{\pi^{3}(1-x_{B}){\langle H\rangle}^{2}_{0}} \cr
\cdot (1-6x_{\beta}/x_{B})
\times\left\{ \frac{1}{4d_{0}}\left [\frac{1}{3}(1-x_{H})\left (1-
5\frac{x_{\beta}}{d_{0}}\right )+\frac{5}{12}x_{B}\left (1+\frac{8x_{\beta}}
{5d_{0}}\right )+ \frac{1}{4}x_{B} \right.\right. \cr
\left.\left. -5x_{\beta}-\frac{1}{3}(x_{H}-x_{B})^{2}\left
(1+4\frac
{x_{\beta}}{d_{0}}\right )\right ]+\frac{1-6x_{\beta}/x_{B}}{1-x_{B}}
\right \} ,
\end{eqnarray}
where $r_{B}\simeq (2m/M_{Z_{2}})^2$, $x_{H}\equiv {(m_{H}/M_{Z_{2}})}^{2}$
$x_{\beta}\equiv\beta/M^{2}_{Z_{2}}$, $d_{0}\simeq\frac{1}{2}
(1+x_{H}-x_{B})$, $\cos\vartheta\equiv (\epsilon\cdot\epsilon_{Z_{2}})$
for $\epsilon^{\mu}$ and $\epsilon^{\mu}_{Z_{2}}$ being
the polarization four-vectors of $B$ and $Z_{2}$ state, respectively.\\
The total relative width $R(Z_{2}\rightarrow H \{\bar{b}b\}_{s=1}/\bar{b}b)$
(\ref{e1}), derived from Eqs. (\ref{e6}) and (\ref{e9}) in the case when $B$
is
composed of $\bar{b}b$ but for the $\bar{b}b$ DY-type normalization, is
presented
in Table 1 as a function of the $H$ boson mass $m_{H}$ via the ratio $x_{H}$.
$~\\ [3 mm]$
{\bf $\underline {Table 1}$ } The values of
$R(Z_{2}\rightarrow H \{\bar{b}b\}_{s=1}/\bar{b}b)\times 10^{10}$ for various
embedding scales $M_{Z_{2}}$ and Higgs boson masses $m_{H}$ via the ratio
$x_{H}={(m_{H}/M_{Z_{2}})}^{2}$.\\
\begin{center}
\begin{tabular}{ c c c c c c c  } \hline
\multicolumn{7}{c}
{$x_{H}$} \\[0.5 mm]
\hline
$M_{Z_{2}}$\,(TeV) & 0 & 0.2 & 0.4 & 0.6 & 0.8 & 0.9  \\[0.5 mm]
\hline
$0.2$ & 2.60 & 2.00 & 1.40 & 0.90 & 0.43 & 0.21  \\[0.5 mm]
$0.3$ & 1.20 & 0.90 & 0.62 & 0.40 & 0.19 & 0.09  \\[0.5 mm]
$0.5$ & 0.42 & 0.32 & 0.23 & 0.15 & 0.07 & 0.03  \\[0.5 mm]
$0.7$ & 0.21 & 0.16 & 0.11 & 0.07 & 0.04 & 0.02  \\[0.5 mm]
\hline
\end{tabular}
\end{center}
To be definite we have considered four values of $M_{Z_{2}}$=0.2, 0.3, 0.5
and 0.7 TeV. As can be seen, the distribution is very steeply peaked towards
low $H$ boson masses and drops to zero at high mass end. In fact, the results
are valid for any masses by simply rescaling the ratios $x_{H,B,\beta}$.
To be understood precisely, one has to note the following:
firstly, the $B$ state is treated relativistically (see (\ref{e8})) and in
the zero binding energy $m_{B}\simeq$ 2m;
secondly, gluon corrections to the process have not been included. For a
heavy
$B$ state such as $\{\bar{b}b\}_{s=1}$ or $\{\bar{t}t\}_{s=1}$, both of
these approximations should be accepted. For a light spin-1 $B$ state
(heavier
Higgs boson), the results can only be taken as a guide of an order of
magnitude of the rates.

The only interesting point has been omitted from our consideration,
namely, the $Z^{\prime}$-decays via spin-1/2 exotic quarks ($h$) with Higgs
boson emission. The exotic quarkonium $\{\bar{h}h\}$ and open flavor
$\{\bar{Q}h\}$ bound states of the exotic $h$-quark production are under
consideration, while they have gained attention in many papers. These
bound states, eg., $\{\bar{h}Q\}$, can be formed since the spectator decays
of a heavy quark constituent $h\rightarrow Q+H$ or $h\rightarrow Q+W(Z)$ are
expected to be suppressed due to small mixing of exotic-SM constituents.
The model presented in this letter is an instructive
one to study and discover extra gauge bosons at LHC and NLC .


\begin {thebibliography}{99}
\bibitem{1}
G. Altarelli et al., Phys. Lett. B 375 (1996) 292.
\bibitem{2}
V. Barger, K. Cheung and P. Langacker, Phys. Lett. B 381 (1996) 226.
\bibitem{3}
e.g. J.L. Hewett, T. Rizzo, Phys. Rep. 183 (1989) 193.
\bibitem{4}
R.W. Robinett, J.L. Rosner, Phys. Rev. D 25 (1982) 3036; P. Langacker,
R.W. Robinett, J.L. Rosner, Phys. Rev. D 30 (1984) 1470;
P. Langacker and M. Luo, Phys. Rev. D 45 (1992) 278.
\bibitem{5}
V. Barger and R.J.N. Philips, Collider Physics (Addison-Wesley, 1987).
\bibitem{6}
G.A. Kozlov, Int. Journ. of Mod. Phys. A 7 (1992) 1935.
\bibitem{7}
S. Ishida, Prog. Theor. Phys. 46 (1971) 1570, 1950; S. Ishida and M. Oda,
Prog. Theor. Phys. 89 (1993) 1033.
\bibitem{8}
V. Bargman, E. Wigner, Proc. Am. Acad. Sci. 34 (1948) 211.

\end{thebibliography}
\end{document}